# On the tractability of the maximum clique problem

R. Dharmarajan[1] and D. Ramachandran[2]

[1]Niels Abel Foundation, Palakkad 678011, Kerala, India.

[1]mathnafrd@gmail.com and [2]claudebergedr@gmail.com

[1]Corresponding author

**Abstract**

The maximum clique problem is a classical NP-complete problem in graph theory and has important applications in many domains. In this paper we show, in a partially non-constructive way, the existence of an exact polynomial-time algorithm for this problem. We outline the algorithm in pseudo-code style. Then we prove its exactness and efficiency by analysis.

**Keywords:** Maximum clique; clique number; induced subgraph; algorithm; polynomial-time.

**Mathematics Subject Classification:** 05C69

## 1. Introduction

The maximum clique problem (MCP) asks for a clique of the largest possible size in a given graph G. Such a clique is called a *maximum clique* of *G*, and the size of any maximum clique of G is the *clique number* (denoted by $\omega(G)$ ) of *G*. The MCP is an NP-complete problem [11]. Since it is one among a number of computational problems that are NP-hard [8], no polynomial-time exact algorithm to solve it is expected to be developed. Nevertheless, it is worthwhile to attempt algorithms for the MCP because it has a wide variety of important applications in fields such as social networking, bioinformatics, document clustering, computer vision, image processing and pattern recognition [1,7,15,21].

There are exact algorithms, approximation algorithms and heuristics for the MCP, the latter two types constituting the class of non-exact MCP algorithms. Informative surveys and reviews on algorithms for the MCP are in [1,4,16,18,25]. Exact algorithms report $\omega(G)$ (*G* being the input graph, or the problem instance) and a maximum clique of *G*. A few outstanding exact algorithms for the MCP are by Harary and Ross [10], Tarjan [19], Carraghan and Pardalos [5] and Ostergaard [14]. Tarjan and Trojanowski [20], Robson [17] and Bourgeois et al [2] gave exact algorithms for the maximum independent set problem which is an NP-complete problem computationally equivalent to the MCP [11,14]. But all the known exact algorithms for the MCP (or any of its equivalent problems) run only in exponential time, and so are not fast in solving practical instances of large sizes.

Non-exact algorithms can run faster than exact ones and can return cliques of sizes approaching (or perhaps equal to) the clique number of the input instance. Many popular non-exact MCP algorithms [1,15,21,18,13,22,23,24] are based on the branch-and-bound approach. Approximation algorithms come with a provable guarantee that the optimal solution is always within a multiplicative factor of the reported solution. Heuristics have no such guarantees. Non-exact algorithms can be of interest in practical applications even though none of the output cliques may be a maximum clique [1]. But such algorithms do not conclusively say anything on the gap between the reported largest clique size and the true clique number of the input graph.

Though MCP algorithms are accompanied by experimental reports and discussions on their performances, very little analysis is given in support of such reports [4]. Consequently, for any algorithm, there seems to be a considerable gap between its reported capabilities and its worst-case performance.

In this article we outline an algorithm (named *ωMAX*) for the MCP. Then, in a partially non-constructive way [3], we follow up with analysis that culminates in proving the exactness and the efficiency of the algorithm.

In section 2 we give relevant definitions (and notation) from graph theory. In section 3 we establish the preliminary results essential to our proposed algorithm *ωMAX*. In section 4 we outline the algorithm *ωMAX* in pseudo-code style. In section 5 we give theory relevant to the running of *ωMAX* and prove its exactness. In section 6 we show our algorithm is of polynomial-time complexity.

## 2. Basic concepts

Let $V$ be a finite nonempty set. The *cardinality* (or, *size*) of $V$ is denoted by $|V|$, and is the number of elements in $V$. The *power set* of $V$ is denoted by $2^V$, and is the set of all the subsets of $V$ including the empty set $\phi$. The set of all nonempty subsets of $V$ is denoted by $2^V*$ - i.e. $2^V = 2^V - \{\phi\}$.

An *undirected simple graph* is an ordered pair $G = (V, E)$ where $V$ is a finite set and $E \subseteq 2^V*$ such that (i) $\cup_{X \in E} X \subseteq V$, and (ii) $|X| \leq 2$ for each $X \in E$. The sets $V$ and $E$ are, respectively, the *vertex set* and the *edge set* of $G$. Each element of $V$ is a *vertex* of $G$ and each member of $E$ is an *edge* of $G$. The integers $|V|$ and $|E|$ are, respectively, the *order* (= the number of vertices) and the number of edges of $G$. The order of $G$ is also denoted by $|G|$. A *loop* is an edge $X$ with $|X| = 1$. G is *loop-free* if $|X| = 2$ for each $X \in E$. If $G$ is loop-free and $\{x, y\}$ is an edge in $G$ then $x$ and $y$ are the *end points* (or, *ends*) of this edge.

Throughout this paper, if $G = (V, E)$, then the expressions $x \in V$ and $x \in G$ will both mean $x$ is a vertex of $G$; similarly, the expressions $\{x, y\} \in E$ and $\{x, y\} \in G$ will both mean $\{x, y\}$ is an edge of $G$. Further, the term graph will mean an undirected simple loop-free graph.

Let $G = (V, E)$ be a graph. Two distinct vertices $x$ and $y$ of $G$ are *adjacent* in $G$ if $\{x, y\} \in E$. If $x$ and $y$ are adjacent in $G$ then each of $x$ and $y$ is a *neighbour* of the other in $G$. For $x \in V$, the set $N(x)$ consisting of all the neighbours of $x$ in $G$ is the *neighbourhood* of $x$ in $G$. The *degree* of $x$ in $G$ is denoted by $dx$ or by $dx(G)$, and is defined as $dx = |N(x)|$. A vertex $y$ of $G$ is *isolated* in $G$ if $dy = 0$. G is *null* if $dx = 0$ for every $x \in V$.

Let $G_1 = (V_1, E_1)$ and $G_2 = (V_2, E_2)$ be graphs. Then $G_1$ is *isomorphic* to $G_2$ if there is a bijective map $f: V_1 \to V_2$ such that for each pair $x$ and $y$ of vertices of $G_1$, $\{x, y\} \in E_1$ if and only if $\{f(x), f(y)\} \in E_2$.

In the following definitions, assume $G = (V, E)$.

A *subgraph* of $G$ is a graph $J = (W, F)$ such that: (i) $W \subseteq V$, (ii) $F \subseteq E$ and (iii) each edge of $J$ has the same end points in $J$ as in $G$. A subgraph $J = (W, F)$ is a *proper subgraph* of $G$ if either $W \neq V$ or $F \neq E$. If $A \subseteq V$ then the *subgraph induced by A* is the subgraph $G[A] = (A, E[A])$ where $E[A]$ is the set of all those edges $\{x, y\} \in E$ such that $x \in A$ and $y \in A$. In particular, if $a \in V$ then the subgraph induced by $V - \{a\}$ will be denoted by $G - a$.

$G$ is *complete* if all of its vertices are pairwise adjacent - i.e. $\{x, y\} \in E$ whenever $x, y \in V$ and $x \neq y$. A *clique* of $G$ is a nonempty set $M \subseteq V$ such that $G[M]$ is complete - i.e. $\{x, y\} \in E$ whenever $x$ and $y$ are distinct elements of $M$. $M$ is a *maximal clique* of $G$ if (i) $M$ is a clique of $G$ and (ii) $M$ is not a proper subset of any clique of $G$. $M$ is a *maximum clique* of $G$ if (i) $M$ is a clique of $G$ and (ii) $|M| \geq |S|$ for every clique $S$ of $G$.

A graph has a maximum clique though such a clique is not necessarily unique. Obviously, if $M_1$ and $M_2$ are maximum cliques of $G$ then $|M_1| = |M_2|$. If $M$ is a maximum clique of $G$ then the positive integer $|M|$ is the *clique number* ($\omega(G)$) of $G$. If $G$ is null then $\omega(G) = 1$.

## 3. Preliminaries

Throughout this section, $G = (V, E)$ and $|G| \geq 2$ are assumed.

**Proposition 3.1.** *Let W be a nonempty proper subset of V. If M is a clique of G such that $M \subseteq W$ then M is a clique of the induced subgraph $G[W]$ of G.*
**Proof.** Let $x$ and $y$ be distinct vertices in $M$. Then the edge $\{x, y\}$ is in $G[W]$ because the end points of this edge ($x$ and $y$) are in $W$. ∎

**Corollary 3.2.** *Let $a \in V$ and M be a clique of G such that $a \notin M$. Then M is a clique of the subgraph $G - a$.*
**Proof.** The graph $G - a$ is the induced subgraph $G[W]$ where $W = V - \{a\}$. Also, $M \subseteq W$. The conclusion now follows from proposition 3.1. ∎

**Proposition 3.3.** $\omega(G) \geq \omega(G - a)$ *for every $a \in V$.*
**Proof.** Every clique of $G - a$ is a clique of $G$. ∎

**Proposition 3.4.** *Suppose $x$ and $y$ are any two distinct non-adjacent vertices of G. Then either $\omega(G) = \omega(G - x)$ or $\omega(G) = \omega(G - y)$.*
**Proof.** Let $\omega(G) = r$, $\omega(G - x) = p$ and $\omega(G - y) = q$. By proposition 3.3, $r \geq p$ and $r \geq q$. We assert that either $r = p$ or $r = q$. Let $M$ be any maximum clique of $G$. Then $|M| = r$. Since $x$ and $y$ are non-adjacent in $G$, either $x \notin M$ or $y \notin M$. Here we invoke Corollary 3.2. If $x \notin M$ then $M$ is a clique of $G - x$ and so $|M| \leq p$. If $y \notin M$ then $M$ is a clique of $G - y$ and so $|M| \leq q$. So either $r \leq p$ or $r \leq q$, whence either $r = p$ or $r = q$. ∎

**Corollary 3.5.** *Suppose G is not complete. Then $\omega(G) = \omega(G - a)$ for some vertex a of G.*

**Corollary 3.6.** *If x and y are distinct non-adjacent vertices of G then $\omega(G) = \omega(G - x) \vee \omega(G - y)$, where $r_1 \vee r_2$ denotes the larger of two given real numbers $r_1$ and $r_2$.*

**Corollary 3.7.** *If G is not complete then for some vertex a of G, every maximum clique of G − a is also a maximum clique of G.*

**Proposition 3.8.** *Let M be a clique of G. Then M is a maximal clique if and only if to each $x \in V - M$ there exists $y \in M$ such that x and y are not adjacent.*
**Proof.** ($\rightarrow$) Assume M is a maximal clique. Let $x \in V - M$ be given. If x were adjacent to each vertex in M then $M \cup \{x\}$ would be a clique in G, contradicting the maximality of M.
($\leftarrow$) If M is not maximal, then $M \cup \{x \setminus\}$ is a clique of G for some $x \in V - M$. So x is adjacent to every vertex in M. ∎

**Proposition 3.9.** *Let W be a nonempty proper subset of V and let $J = (W, F)$ be the subgraph (of G) induced by W. Let $x \in V - W$ and suppose x is adjacent to every vertex of J. Let H be the subgraph induced by $W \cup \{x\}$. Then:*
*(i) $\omega(H) = \omega(J) + 1$.*
*(ii) Suppose $M \subseteq W$. Then $M \cup \{x\}$ is a maximum clique of H if and only if M is a maximum clique of J.*
**Proof.** (i) Obviously x is adjacent to every vertex of J. Let $\omega(J) = p$ and M be a maximum clique of J. Then $|M| = p$, $x \notin M$ and $M \cup \{x\}$ is a clique of H. So $\omega(H) \geq p + 1$. If H had a clique of size $p + 2$ then $H - x$ would have a clique of size at least $p + 1$. This, in view of $H - x$ being isomorphic to J, contradicts $\omega(J) = p$. This proves (i).
(ii) Clearly $x \notin M$. Assume $M \cup \{x\}$ is a maximum clique of H. Let $|M \cup \{x\}| = p + 1$. Then $\omega(H) = i + 1$. Also, M is a clique of J. If J had a clique S such that $|S| > |M|$ then $S \cup \{x\}$ would be a clique of H owing to x being adjacent to every vertex of J. This would mean $\omega(H) \geq |S \cup \{x\}| > p + 1$, patently contradicting $\omega(H) = p + 1$. Hence M is a maximum clique of J.
Conversely, suppose M is a maximum clique of J and $|M| = p$. Then clearly $M \cup \{x\}$ is a clique of H. If H were to have a clique S of size exceeding $p + 1$ then $S - \{x\}$ would be a clique of J of size exceeding p, an impossibility. Further, $|M \cup \{x\}| = p + 1$. ∎

**Proposition 3.10.** *Upto isomorphism, there is only one graph $G = (V, E)$ such that $|V| = 3$ and $|E| = 1$.*
**Proof.** Let $V = \{x, y, z\}$ and E consist of the edge $\{x, y\}$. Let $J = (W, F)$ be any graph with order 3 and exactly one edge. Let $W = \{a, b, c\}$ and F consist of the edge $\{a, b\}$. Then the map $h: G \rightarrow J$ defined by $h = \{(x, a), (y, b), (z, c)\}$ is an isomorphism of G onto J. ∎

**Proposition 3.11.** *Upto isomorphism, there is only one graph $G = (V, E)$ such that $|V| = 3$ and $|E| = 2$.*
**Proof.** Let $V = \{x, y, z\}$ and E consist of the two edges $\{x, y\}$ and $\{x, z\}$. Let $J = (W, F)$ be any graph with order 3 and exactly two edges. Let $W = \{a, b, c\}$ and F consist of the two edges {a, b} and {a, c}. Then the map h: G → J defined by h = {(x, a), (y, b), (z, c)} is an isomorphism of G onto J. ∎

# 4. The proposed algorithm ω MAX

The principal idea behind this algorithm is Proposition 3.4: if G is not complete then $\omega(G) = \omega(J)$ for some proper subgraph J of G.

**Input:** The vertex set V and the edge set E of a graph G = (V, E).

### Pre-processing
(i) Compute the adjacency list of G
(ii) Compute the adjacency matrix of G
(iii) Go to the main algorithm

### The main algorithm

BEGIN with (1)
(1) (i) Compute $n = |V|$ and $e = |E|$ ; (ii) order the vertices of G as $x_1, \ldots, x_n$
where $dx_j \geq dx_{j+1}$ for $j = 1, \ldots, n − 1$ and (iii) $Ver = [x_1, \ldots, x_n]$ ; go to (2)
(2) If $e = n(n − 1) / 2$ then **output:** (i) $\omega(G) = n$ and (ii) the only maximum clique of G is V and END, else go to (3)
(3) If $e = 0$ then **output:** (i) $\omega(G) = 1$ and (ii) the only maximum cliques of G are the singleton subsets of V and END, else go to (4)
(4) LC(STORED) = $\{x_1\}$ ; ω(STORED) = 1 and $r = 2$; go to (5)
(5) $W = [Ver(1), \ldots, Ver(r)]$ ; go to (6)
(6) If the last element, say z, of W is adjacent to every other element of W
then ω(STORED) ← ω(STORED) + 1 and LC(STORED) ← LC(STORED) ∪ {z}
and go to (7) else go to (8)
(7) $r ← r + 1$ and return to (5)
(8) Let m be the largest index ($\leq r − 1$) such that $x_m \in W$ and $x_m$ is not adjacent to the last element of W; go to (9)
(9) (i) Let $S = [W(1), \ldots, W(m − 1), W(m + 1), \ldots, W(r)]$ ; go to (10)
(10) $j = 1$; go to (11)
(11) $IP = S$; $OP = \phi$; go to (12)
(12) $Lead = IP(1)$ ; $OP ← [OP, Lead]$; $RevIP = \phi$; go to (13)
(13) $k = |IP|$ ; go to (14)
(14) if $k > 1$ got to (15), else go to (16);
(15) for $a = 2$ to $k$
  if $IP(a)$ is adjacent to Lead then $RevIP ← [RevIP, a]$
  else end
 end for; go to (16)
(16) if $RevIP = \phi$ then go to (18) else go to (17);
(17) $IP ← RevIP$ and go to (12)

(18) ω(CURRENT) = |OP| ; go to (19)

(19) If ω(CURRENT) > ω(STORED)

then ω(STORED) ← ω(CURRENT) and LC(STORED) ← OP

else end; go to (20)

(20) $j \leftarrow j + 1$; go to (21)

(21) If $j \leq |S|$ then $S \leftarrow [S(2), \ldots, S(r-1), S(1)]$ and go to (11)

else $r \leftarrow r + 1$ and go to (22)

(22) if $r \leq n$ go to (5) else go to (23)

(23) **OUTPUT:** (i) ω(G) = ω(STORED) and (ii) LC(STORED) is a maximum clique of G

END

**Remark.** The notation $S \leftarrow [S(2), \ldots, S(r-1), S(1)]$ in step (21) of the pseudo-code is a cyclic permutation of the current $S$. It results in the current $S$ being replaced by $\sigma(S)$ where $\sigma$ is the permutation defined on the set $\{1, \ldots, k\}$ by $\sigma(i) = i + 1$ for $i = 1, \ldots, k - 1$ and $\sigma(k) = 1$.

## 5. Proof of the exactness of ω MAX

A few definitions and notations are needed to discuss the feasibility and capabilities of ωMAX. Let $X$ be a nonempty finite set and $|X| = n$. An ordered set (or, an ordered $r$-set) over $X$ is a 1 x $r$ matrix $P = [x_1, \ldots, x_r]$ where (i) $0 \leq r \leq n$, (ii) $x_j \in X$ for $j = 1, \ldots, r$ and (iii) if $x_j, x_k \in X$ with $1 \leq j < k \leq r$ then $x_j \neq x_k$. The set $X$ is the base set for $P$. The integer $r$ is the *cardinality* (or, *size*) of $P$. As with unordered sets, $|P|$ denotes the size of the ordered set $P$. If $r = 0$ then $P$ is the *empty ordered set* ($\phi$) over $X$. An ordered $r$-set will be referred to as an ordered set unless the mention of $r$ is warranted. An ordered set over $X$ will be referred to as an ordered set if $X$ is understood from the context. The empty ordered set will be called the empty set. If $P = [x_1, \ldots, x_r]$ is an ordered set over $X$ and $x \in X$ then $x \in P$ if and only if $x = x_j$ for some $j$ with $1 \leq j \leq |P|$. For $j \in \{1, \ldots, r\}$, $P(j)$ will denote the $j^{th}$ element of $P$ (which is $x_j$).

Let $P_1$ and $P_2$ be two ordered sets over $X$. Write $P_1 = [x_1, \ldots, x_k]$ and $P_2 = [y_1, \ldots, y_r]$. Then $P_1 = P_2$ if and only if (i) $k = r$ and (ii) $x_j = y_j$ for $j = 1, \ldots, k$. If $P = [x_1, \ldots, x_r]$ is an ordered set over $X$, $y \in X$ and $y \notin P$ then the *augmentation* of $P$ by $y$ (or, the $y$-*augumentation* of $P$) is denoted by $[P, y]$ and is defined to be the ordered set $[x_1, \ldots, x_r, y]$.

In the algorithm ωMAX, the first iteration begins when $r$ is assigned the value 2 in step (4) of the pseudo-code. It ends when $r \leftarrow r + 1$ is executed, which happens in step (7) or step (21) according as the control is directed to. This is also where the second iteration begins. In general, the $i^{th}$ iteration begins when step (7) or (21) does the assignment $r \leftarrow r + 1$. By (4) and (22) it is clear that there are exactly $n - 1$ iterations (corresponding to $r = 2$ through $n$) if the input graph has order $n$. From the beginning of an iteration to its end, the value of r is constant.

In each iteration, there are sub-iterations corresponding to $j = 1$ through $|S|$. It is clear from steps (5) and (9) that $|S| = r - 1$. Hence if an iteration corresponds to $r = k$ then this iteration consists of $k - 1$ sub-iterations. The total number of sub-iterations in the algorithm is, then, $1 + \ldots + n - 1$ which is $n(n-1)/2$.

**Proposition 5.1.** *The sets Ver, W, S, IP, RevIP and OP are all ordered sets over the base set V, the vertex set of the instance graph G.*
**Proof.** Ver = $[x_1, \ldots, x_n]$ is an ordered set over V, from (1)(iii) of the pseudo-code of $\omega MAX$. So are W and S, from (5) and (9), respectively. (11) makes it clear that IP is an ordered set over V. (12) and (15) show, respectively, that OP and RevIP are ordered sets over V. ∎

In $\omega MAX$, if A is an ordered set and m is a positive integer, then A(m) will denote the $m^{th}$ element of A. For instance, W(m − 1) in step 9 is the $(m - 1)^{th}$ element of W, and IP(1) in step 12 is the first element of IP.

**Proposition 5.2.** *The algorithm $\omega MAX$ terminates in a finite number of computations.*
**Proof.** The pre-processing phase of the $\omega MAX$ terminates in finitely many computations because V and E are finite sets. Let n be the order of the input graph G. If G is null or complete, then the running of $\omega MAX$ begins with step (1) and terminates with either step (2) or step (3). Each of these steps executes only a finite number of computations. So suppose G is neither null nor complete.

Steps (4) through (9) clearly involve only finitely many computations. So do (20) through (23).

The first iteration begins when $r = 2$ and the last iteration begins when $r = n$. For each $r = 2, \ldots, n$, the first sub-iteration begins when $j = 1$ and the last subiteration begins when $j = |S| = r − 1$. For each j, the loop in (15) is executed $k − 1$ times for $k > 1$, where $k = |IP|$. But then by (15) and (17), the value of k decreases by at least 1 every time the control returns to (12), and so $k = 1$ happens in a finite number of computations. Hence (15) is done only finitely many times. Also, the steps (11) through (14) and (16) through (19) all depend on $|IP|$. Since RevIP = $\phi$ must happen when $k = 1$, it follows that (11) through (14) and (16) through (19) all involve only a finite number of computations. Consequently, $\omega MAX$ executes only finitely many computations for each $r = 2, \ldots, n$.

Next, by (21) it is clear that $r = n + 1$ happens after finitely many iterations. By (22) and (23), it is clear that the algorithm terminates when $r = n + 1$. ∎

The ordered set OP obained at the end of a sub-iteration is called the *ordered output set* from this sub-iteration, or, simply, an *ordered output set*, if the ordinal of the sub-iteration need not be mentioned. An ordered output set will be denoted by OOS.

**Proposition 5.3.** *If OP = $[y_1, \ldots, y_k]$ is an OOS, then the vertex $y_j$ is adjacent to each of $y_{j+1}$ through $y_k$ for each $j = 1, \setminus \ldots, k − 1$.*
**Proof.** Suppose there were $y_p, y_t \in OP$ ($1 \leq p < t \leq k$) with $y_p$ not adjacent to $y_t$. Since $y_p$ is in OP, it happened that Lead = $y_p$ at some point of the sub-iteration under consideration. Then when (15) is executed in this sub-iteration, $y_t$ would not be included in RevIP. Then $y_t$ would not be in IP at any subsequent point in this sub-iteration, and so $y_t$ would not be in OP at the end of this sub-iteration. But this contradicts $y_t \in OP$. ∎

**Proposition 5.4.** *Let OP= $[y_1, \ldots, y_t]$ be an OOS. If $x \in V$ and $x \notin OP$, then $x \notin N(y_j)$ for some $y_j$ in OP.*

**Proof.** Note that $|OP| = t$. Clearly $x$ was not *Lead* at any point - else $x$ would be in *OP* by dint of (12). Next, if it were true that $x \in N(y_j)$ for every $y_j \in OP$, then when *Lead* = $y_t$, we would subsequently have $x \in RevIP$, and so $RevIP \neq \phi$. Then there would be a subsequent computation in which *Lead* = $y_q$ for some $y_q \in RevIP$ and $q > t$, with $y_q \neq x$. This would entail $y_q \in OP$ at the end of the iteration, thereby giving $y_j \in OP$ for $j = 1, \ldots, t, q$, leading to $|OP| > t$, a patent contradiction. ∎

**Corollary 5.5.** *Let OP= $[y_1, \ldots, y_t]$ be an OOS at the end of a sub-iteration. If $x \in V$ and $x \notin OP$ then for some $z \in OP$, $x$ was not augmented to RevIP at some point during the sub-iteration when Lead = z.*

**Proof.** If the conclusion were not true for any $z \in OP$ at the end of the sub-iteration, then $x$ would be a neighbour of each vertex of *OP*. But then $x$ would consequently be in *OP*, contradicting $x \notin OP$. ∎

**Proposition 5.6.** *Let OP= $[y_1, \ldots, y_t]$ be an OOS. Then then the set $M = \{y_1, \ldots, y_t\}$ is a maximal clique of G.*

**Proof.** If $y_i$ and $y_j$ are in *OP* where $i \neq j$, then $y_i$ is adjacent to $y_j$ by proposition 5.1. So *M* is a clique of *G*. Next, let $x \in V - M$ be given. Let *OP* be the *OOS* at the end of a sub-iteration. Since $x \notin OP$, $x$ was excluded form *RevIP* by some vertex $z$ in a computation subsequent to *Lead* = $z$ during the sub-iteration. Clearly, then, $z$ and $x$ are not neighbours. Further, since *Lead* = $z$ during this sub-iteration, we have that $z \in OP$ at the end of this sub-iteration, whence $z = y_p$ for some $p \in \{1, \ldots, t\}$. So $x$ is not adjacent to the vertex in $y_p \in M$. By dint of proposition 3.8, *M* is a maximal clique of *G*. ∎

**Proposition 5.7.** *If $G = (V, E)$ is complete or null, then ωMAX returns ω(G) and a maximum clique in G.*

**Proof.** Let $n = |V|$ and $e = |E|$. Suppose *G* is complete. Then $e = |E| = n(n-1)/2$. The algorithm checks this to be true - from steps (1) and (2) - and reports: (i) *ω(G)* = n and (ii) *V* is the only maximum clique of *G*.

Suppose *G* is null. Then $e = 0$. The algorithm checks this to be true - from steps (1) and (3) - and reports: (i) *ω(G)* = 1 and (ii) the only maximum cliques of *G* are the singleton subsets of *V*. ∎

**Proposition 5.8.** *Let $G = (V, E)$ be of order 3. The algorithm ωMAX returns the clique number of G and a maximum clique of G.*

**Proof.** Let $V = \{x, y, z\}$. If G is complete then *ωMAX* runs on *G* as follows:
(1) $n = 3$; $e = 3$
(2) $e = n(n-1)/2$; so report: (i) *ω(G)* = 3 and the only maximum clique of *G* is *V*.

If *G* is null then *ωMAX* runs on *G* as follows:
(1) $n = 3$; $e = 0$

(2) $e = 0$, so report: (i) $\omega(G) = 1$ and the maximum cliques of $G$ are the singleton subsets of $V$.

If $G$ is neither complete nor null, then $G$ has at least one edge and at most two edges.

**Case 1.** Suppose $E$ is the set consisting of one edge - say, $\{x, y\}$. Here $Ver = [x, y, z]$ in accordance with (ii) and (iii) of step (1) of the pseudo-code, since $dx = dy = 1$ and $dz = 0$. Then $\omega MAX$ runs on $G$ as follows (with each right arrow denoting the passing from one computation to the logical next):

$n = 3 \to e = 1 \to e \neq 0 \to e \neq n(n-1)/2 \to LC(STORED) = \{x\} \to \omega(STORED) = 1 \to r = 2 \to W = [x, y] \to \omega(STORED) = 2 \to LC(STORED) = \{x, y\} \to r = 3 \to W = [x, y, z] \to S = [x, z] \to j = 1 \to IP = [x, z] \to OP = \phi \to Lead = x \to OP = [x] \to RevIP = \phi \to k = 2 \to \omega(CURRENT) = 1 \to j = 2 \to S = [z, x] \to IP = [z, x] \to OP = \phi \to Lead = z \to OP = [z] \to RevIP = \phi \to k = 2 \to \omega(CURRENT) = 1 \to j = 3 \to r = 4 \to \omega(G) = 2 \to$ a maximum clique of $G$ is $\{x, y\} \to END$.

**Case 2.** Suppose $E$ consists of two edges - say, $\{x, y\}$ and $\{x, z\}$. Here $Ver = [x, y, z]$ since $dx = 2$ and $dy = dz = 1$. Then $\omega MAX$ runs on $G$ as follows:

$n = 3 \to e = 2 \to e \neq 0 \to e \neq n(n-1)/2 \to LC(STORED) = \{x\} \to \omega(STORED) = 1 \to r = 2 \to W = [x, y] \to \omega(STORED) = 2 \to LC(STORED) = \{x, y\} \to r = 3 \to W = [x, y, z] \to S = [x, z] \to j = 1 \to IP = [x, z] \to OP = \phi \to Lead = x \to OP = [x] \to RevIP = \phi \to k = 2 \to RevIP = [z] \to IP = [z] \to Lead = z \to OP = [x, z] \to RevIP = \phi \to k = 1 \to \omega(CURRENT) = 2 \to j = 2 \to S = [z, x] \to IP = [z, x] \to OP = \phi \to Lead = z \to OP = [z] \to RevIP = \phi \to k = 2 \to RevIP = [x] \to IP = [x] \to Lead = x \to OP = [z, x] \to RevIP = \phi \to k = 1 \to \omega(CURRENT) = 2 \to j = 3 \to r = 4 \to \omega(G) = 2 \to$ a maximum clique of $G$ is $\{x, y\} \to END$

**OUTPUT:** $\omega(G) = 2$ and $\{x, y\}$ is a maximum clique of $G$.

Thus, when $G$ has order $n = 3$, $\omega MAX$ returns $\omega(G)$ and a maximum clique of $G$. ∎

**Proposition 5.9.** *Let $G = (V, E)$ be of order $n$. Then there is a linear ordering of vertices of $G$ for which $\omega MAX$ returns $\omega(G)$ and a maximum clique of $G$, by returning $\omega(J)$ and a maximum clique of $J$ for some subgraph $J$ of $G$ such that $\omega(G) = \omega(J)$. In other words, $\omega MAX$ converges to a desired output, which is a pair $(\omega(G), M)$ where $M$ is a maximum clique of $G$.*

**Proof.** Let the proposition be called $P(n)$. The proof is by induction on $n$. The proof for $n = 3$ was done by proposition 5.8.

Assume $P(k)$ is true. Next, suppose $n = k + 1$. If $G$ is complete then, by proposition 5.7, $\omega MAX$ returns (i) $\omega(G) = k + 1$ and (ii) $V$ is the only maximum clique of $G$. If $G$ is null then, again by proposition 5.7, the algorithm returns (i) $\omega(G) = 1$ and (ii) the only maximum cliques of $G$ are the singleton subsets of $V$. In both these cases the subgraph $J$ in the statement of the current proposition is $G$. So $P(k + 1)$ is true when $G$ is complete or null.

Suppose $G$ neither complete nor null. Then $G$ has two vertices - say, $a$ and $b$ - that are not neighbours in $G$. Clearly $G - a$ and $G - b$ have order $k$. By the induction hypothesis, there is a linear vertex ordering in $G - a$ and one in $G - b$ for which $\omega MAX$ returns the following:

(i) $\omega(G - a) = r_1$ (say), (ii) a maximum clique $M_1$ of $G - a$, (iii) $\omega(G - b) = r_2$ (say) and (iv) a maximum clique $M_2$ of $G - b$. Here $|M_1| = r_1$ and $|M_2| = r_2$.

Let $r = r_1 \vee r_2$. By corollary 3.6, $\omega(G) = r$. Further, either $M_1$ or $M_2$ is a clique of size $r$ - and hence a maximum clique of $G$. But then $\omega MAX$ has returned $r$ and a clique $M$ ($=M_1$ or $M_2$) that are respectively $\omega(G)$ and a maximum clique of $G$. The required subgraph $J$ in the current proposition is either $G - a$ (if $r = r_1$) or $G - b$ (if $r = r_2$). And the required linear ordering of vertices of $G$ is the one in $G - a$ (if $r = r_1$) or the one in $G - b$ (if $r = r_2$). This completes the induction. ∎

## 6. The worst-case time complexity of $\omega MAX$

The worst-case time complexity of $\omega MAX$ is discussed using the asymptotic growth rate function $O$ (big *oh*). The term time complexity will mean the worst-case one, throughout this section.

**Primitive computational steps.** Throughout this section, by the phrases "(*) is bounded by time $O(n^d)$" and "(*) takes $O(n^d)$ time," we will mean that there are absolute constants $c > 0$ and $d > 0$ so that on every input graph of order $n$, the running time of the process in the place of (*) is bounded by $cn^d$ primitive computational steps ([12], chapter 2). The following are the primitive computational steps in $\omega MAX$:

(p-c 1) Assigning a value to a variable;
(p-c 2) placing a new element at the end of a list of elements;
(p-c 3) reading an element from a list; and
(p-c 4) any of the four fundamental operation on real numbers.

Further, the term "instance" in this section will mean an input graph. For each instance, steps (1) through (4) of $\omega MAX$ are run once, and so is step (23). Steps (5) through (9) are run at most $n$ times each. Steps (10) through (22) are run at most $n^2$ times each. In the algorithm, the positive integer $r$ ranges from 2 to $n$ (the order of the input instance) and each value of $r$ corresponds to an iteration.

The computation of the adjacency matrix is bounded by time $O(n^3)$; so is the computation of the adjacency list. Consequently the pre-processing phase of $\omega MAX$ is bounded by $O(n^3)$ time. In the following analysis of the main algorithm phase of $\omega MAX$, the time complexity of steps (1) through (4) and (23) are for one instance and that of steps (5) through (22) are for an arbitrary iteration.

In step (1) of the pseudo-code (section 4), computing $|V|$ takes $O(n)$ time whereas computing $|E|$ takes $O(n^2)$ time. Next, computing the ordered set *Ver* takes $O(n)$ time. Hence step (1) is bounded by time $O(n^2)$. Step (2) is bounded by time $O(n)$; so is (3). If $G$ is neither complete nor null then the control goes to (4). The assignations $\omega(STORED) = 1$ and $LC(STORED) = \{x_1\}$ require time $O(1)$ each. Hence steps (1) through (4) are bounded by time $O(n^2)$.

Step (5) is bounded by $O(n)$ since it is a finite sequence of primitive computational step of the type (p-c 2) seen above. In step (6), checking if $Ver(r)$ is adjacent to any other element of $W$ can be done in $O(n)$ time using the adjacency matrix. The assignations seen in (6) are bounded by time $O(n)$. (7) obviously requires only constant time. Finding an element $x_m$ as in (8) is done using the adjacency matrix, and so (8) is bounded by O($n^2$). (9) is similar to (5) and so is bounded by O(n). Obviously (10), (11) and (12) require only constant time. (13) is bounded by O(n) since $|IP| \leq n$. (14) is done in constant time.

Whether *a* is adjacent to *Lead* is determined by checking the adjacency matrix for an edge connecting *Lead* and *a*. This can be done in constant time ([12], chapter 3). Further, $k \leq n - 1$. So each time the *for* loop in (15) is executed fully, there are at most $\sum(n - 1)$ computational steps, each of constant time. In each iteration, (15) is run at most *n* times. Hence (15) is bounded by time O($n^3$).

In (16), the algorithm needs to access only the first element of *RevIP* (if there is one). So (16) takes O(1) time. (17) is bounded by time O(n); so is (18). In (19), the logical operation requires constant time, as does the assignation $\omega(STORED) \leftarrow \omega(CURRENT)$. The other assignation *LC(STORED) = OP* is bounded by time O(n). So (19) takes time O(n). It is clear that (20) is done in O(1) time. Next, (21) is bounded by time O(n) whereas (22) and (23) are each bounded by O(1) time.

Hence each iteration is bounded by time O($n^3$). Since there are $n - 1$ iterations for each instance (corresponding to $r = 2$ through *n*), the time complexity of $\omega MAX$ is O($n^4$).

## 7. An example and some comments

Let $G = (V, E)$ be the graph with $V = \{1, 2, 3, 4, 5, 6, 7\}$. The adjacency list of $G$ is:
(i) $N(1) = \{2, 3, 4\}$, (ii) $N(2) = \{1\}$, (iii) $N(3) = \{1\}$, (iv) $N(4) = \{1\}$, (v) $N(5) = \{6, 7\}$, (vi) $N(6) = \{5, 7\}$ and (vii) $N(7) = \{5, 6\}$, where $N(x)$ denotes the neighbourhood (in $G$) of the vertex $x \in G$.

The vertices of $G$ in a non-ascending order of degrees are: 1, 5, 6, 7, 2, 3, 4. In the following steps, the numbering (1) through (9) is not connected with that in the pseudo-code of $\omega MAX$. The arrows ($\rightarrow$) indicate the sequence of computations.

BEGIN with (1)
(1) $n = |V| = 7$, $e = |E| = 6$ and $Ver = [1, 5, 6, 7, 2, 3, 4]$. Go to (2).

(2) $\omega(STORED) = 1$ and $LC(STORED) = \{1\}$. Go to (3).

(3) $r = 2 \rightarrow W = [1, 5] \rightarrow S = [5] \rightarrow j = 1 \rightarrow \omega(CURRENT) = 1 \rightarrow j = 2$. Go to (4).

(4) $r = 3 \rightarrow W = [1, 5, 6] \rightarrow S = [5, 6] \rightarrow j = 1 \rightarrow \omega(CURRENT) = 2 \rightarrow \omega(STORED) = 2 \rightarrow LC(STORED) = \{5, 6\} \rightarrow j = 2 \rightarrow \omega(CURRENT) = 2 \rightarrow j = 3$. Go to (5).

(5) $r = 4 \rightarrow W = [1, 5, 6, 7] \rightarrow S = [5, 6, 7] \rightarrow j = 1 \rightarrow \omega(CURRENT) = 3 \rightarrow \omega(STORED) = 3 \rightarrow LC(STORED) = \{5, 6, 7\} \rightarrow j = 2 \rightarrow \omega(CURRENT) = 3 \rightarrow j = 3 \rightarrow \omega(CURRENT) = 3 \rightarrow j = 4$. Go to (6).

(6) $r = 5 \to W = [1, 5, 6, 7, 2] \to S = [1, 5, 6, 2] \to j = 1 \to \omega(CURRENT) = 2 \to j = 2 \to \omega(CURRENT) = 2 \to j = 3 \to \omega(CURRENT) = 2 \to j = 4 \to \omega(CURRENT) = 2 \to j = 5$. Go to (7).

(7) $r = 6 \to W = [1, 5, 6, 7, 2, 3] \to S = [1, 5, 6, 7, 3] \to j = 1 \to \omega(CURRENT) = 2 \to j = 2 \to \omega(CURRENT) = 3 \to j = 3 \to \omega(CURRENT) = 3 \to j = 4 \to \omega(CURRENT) = 3 \to j = 5 \to \omega(CURRENT) = 2 \to j = 6$. Go to (8).

(8) $r = 7 \to W = [1, 5, 6, 7, 2, 3, 4] \to S = [1, 5, 6, 7, 2, 3] \to j = 1 \to \omega(CURRENT) = 2 \to j = 2 \to \omega(CURRENT) = 3 \to j = 3 \to \omega(CURRENT) = 3 \to j = 4 \to \omega(CURRENT) = 3 \to j = 5 \to \omega(CURRENT) = 2 \to j = 6 \to \omega(CURRENT) = 2 \to j = 7$. Go to (9).

(9) $r = 8 \to \omega(G) = 3$ and $\{5, 6, 7\}$ is a maximum clique of $G$.
END

**Comments on Propositions.** Propositions 3.1 and 3.3 are important to establish proposition 3.4. Proposition 3.4 proves that for a given graph $G = (V, E)$ that is not complete, there is a proper subgraph $J$ such that $\omega(G) = \omega(J)$. This is crucial to $\omega MAX$ because the required subgraph $J$ is either $G - x$ or $G - y$, where $x$ and $y$ are any two non-adjacent vertices of $G$. Inherent in the algorithm is the existence of a linear ordering of the elements of some nonempty subset $W$ of $V$. The $\omega MAX$ returns an optimal solution under this linear ordering of the elements of $W$. Such an ordering is explicit in the proof of proposition 5.8 and implicit in the proof of proposition 5.9.

Proposition 5.9 proves the exactness of $\omega MAX$ - i.e. for a given instance $G$, $\omega MAX$ does return $\omega(G)$ and a maximum clique of $G$. Propositions 5.7 and 5.8 are essential to proposition 5.9. The propositions 5.1 through 5.6 of section 5 prove the feasibility and some capabilities of $\omega MAX$.

**The non-constructive facet of ω MAX.** Since $\omega(G)$ is known for every graph $G$ of order 3, the proof of proposition 5.8 (the base case for our induction) was straightforward.
However, in the proof of proposition 5.9, it is reasoned that there exists an ordering of vertices of $G$ for which $\omega MAX$ returns $\omega(G)$, but such an order is not constructed explicitly here. This linear vertex ordering is the only aspect of $\omega MAX$ that is not explicit, which is the reason we deem $\omega MAX$ partially non-constructive.

# 8. Concluding remarks

We have presented an exact polynomial-time algorithm for determining $\omega(G)$ of a given graph $G$. The independence number and the vertex cover number of a given graph $G$ can be found by applying $\omega MAX$ on the complement graph [15] of $G$. We have not reported any experimentation with $\omega MAX$ because we have analytically proved its exactness and efficiency (in sections 5 and 6, respectively).

To sum up: (i) there exists an exact polynomial-time algorithm for the MCP for all graphs of order 3 (Proposition 5.8) and (ii) if this algorithm returns $\omega(G)$ for all graphs of $G$ order $k$ then the algorithm returns $\omega(G)$ for all graphs $G$ of order $k + 1$ (Proposition 5.9). Hence there exists an exact polynomial-time algorithm ($\omega MAX$) for the MCP for all graphs of order $n \geq 3$. Symbolically: (i) *P(3)* is true and (ii) *P(k + 1)* is true whenever *P(k)* is true; hence *P(n)* is true for all $n \geq 3$.

Thus, the MCP is tractable. For an account of tractability of problems, we refer the reader to [3,6,9,12].

# Acknowledgements

This research was fully supported by Niels Abel Foundation, Palakkad, Kerala State, India. The authors thank Dr. K. Kannan, Professor of Mathematics with SASTRA University (Thanjavur, India) and Dr. V. Swaminathan, Assistant Professor of Mathematics with SASTRA University SRC (Kumbakonam, India) for discussions and their suggestions on analysing algorithms.

# References

[1] I. M. Bomze, M. Budinich, P. M. Pardalos, M.Pelillo, The maximum clique problem, Handbook of Combinatorial Optimization **4** (1999) 1-74. Doi.10.1.1.56.6221

[2] N. Bourgeois, B. Escoffier, V. T. Paschos, J. M. M. van Rooij, Fast Algorithms for max independent set, Algorithmica **62**(1) (2012) 382-415.

[3] D. P. Bovet and P. Crescenzi, Introduction to the theory of complexity, Creative Commons Attribution - NonCommercial 2.5, Citeseer (2006).

[4] R. Carmo and A. Zuge, Branch and bound algorithms for the maximum clique problem under a unified framework. J. Braz. Comput. Soc. **18**(2) (2012) 137-151. Doi: 10.1007/s13173-011-0050-6

[5] R. Carraghan, P. M. Pardalos, An exact algorithm for the maximum clique problem, Oper. Res. Lett. **9**(6) (1990) 375-382.

[6} T. H. Cormen, C. E. Leiserson, R. L. Rivest, C. Stein, Introduction to Algorithms (MIT Press, 2009).

[7] D. Eppstein, M. Loffler, D. Strash, Listing all maximal cliques in sparse graphs in near-optimal time, in Proc. Intl. Symp. on Algorithms and Computation, LNCS **6506** (Springer, 2010), pp. 403-414.

[8] M. R. Garey, D. S. Johnson, Computers and Intractability: A guide to the Theory of NP-completeness, Computers and Intractability **340** (1979).

[9] S. H. Gerez, Algorithms for VLSI Design Automation (John Wiley and Sons, 1999).

[10] F. Harary, I.C. Ross, A Procedure for Clique Detection Using the Group Matrix, Sociometry **20**(3) (1957) 205-215.

[11] R. Karp, Reducibility among Combinatorial Problems, in Complexity of Computer Computations (Springer, 1972), pp. 85-103.

[12] J. Kleinberg, E. Tardos, Algorithm Design (Pearson Education, 2006).

[13] J. Konc, D. Janezic, An improved branch and bound algorithm for the maximum clique problem, MATCH Commun. Math. Comput. Chem. **58**(3) (2007) 569-590.

[14] P. R. J. Ostergard, A fast algorithm for the maximum clique problem, Discrete Appl. Math. **120** (2002) 197-207.

[15] P.M. Pardalos, J. Xue, The maximum clique problem, J. Global Optim. **4**(3) (1994) 301-328.

[16] P. Prosser, Exact Algorithms for Maximum Clique: A Computational Study, Algorithms **5**(4) (2012) 545-587.

[17] J. M. Robson, Algorithms for maximum independent sets, J. Algorithms **7**(3) (1986) 425-440.

[18] K. Singh, A. K. Pandey, Survey of Algorithms on Maximum Clique Problem, Int. Advanced Research J. in Sci., Eng. and Tech. **2**(2) (2015) 15-20.

19] R. E. Tarjan, Finding a maximum clique, Technical Report TR 72 - 123 (1972), Computer Science Department, Cornell University, Ithaca, New York.

[20] R. E. Tarjan, A. E. Trojanowski, Finding a maximum independent set, SIAM J. Comput. **6**(3) (1977) 537-546.

[21] E. Tomita, T. Akutsu, T. Matsunaga, Efficient Algorithms for Finding Maximum and Maximal Cliques: Effective Tools for Bioinformatics, in Biomedical Engineering, Trends in Electronics, Communications and Software (2011) 625-640.

[22] E. Tomita, T. Kameda, An efficient branch-and-bound algorithm for finding a maximum clique with computational experiments, J. Glob. Optim. **37**(1) (2007) 95-111. Doi: 10.1007/s10898-006-9039-7

[23] E. Tomita, T. Seki, An efficient branch-and-bound algorithm for finding a maximum clique, in Discrete Mathematics and Theoretical Computer Science, LNCS **2731** (Springer, 2003), pp. 278-289.

[24] E. Tomita, Y. Sutani, T. Higashi, M. Wakatsuki, A simple and faster branch-and-bound algorithm for finding a maximum clique with computational experiments, IEICE Trans. Inf. \& Syst. **96**(6) (2013) 1286-1298.

[25] Q. Wu, J. Hao, A review on algorithms for maximum clique problems, European J. Oper. Res. **242**(3) (2015) 693-709.

---